\documentclass[11pt, twoside, letter]{llncs}
\usepackage{color}
\usepackage{url}
\usepackage{geometry}
\usepackage{epsfig}
\global\emergencystretch = .9\hsize

\twocolumn

\begin{document}
\title{The topology of covert conflict}
\author{Shishir Nagaraja, Ross Anderson}
\institute{Computer Laboratory\\
JJ Thomson Avenue, Cambridge CB3 0FD, UK\\
{\tt forename.surname @ cl.cam.ac.uk}}
\maketitle

\begin{abstract}
  Often an attacker tries to disconnect a network by destroying nodes
  or edges, while the defender counters using various resilience
  mechanisms. Examples include a music industry body attempting to
  close down a peer-to-peer file-sharing network; medics attempting to
  halt the spread of an infectious disease by selective vaccination;
  and a police agency trying to decapitate a terrorist organisation.
  Albert, Jeong and Barab\'asi famously analysed the static case, and
  showed that vertex\\-order attacks are effective against scale-free
  networks. We extend this work to the dynamic case by developing a
  framework based on evolutionary game theory to explore the
  interaction of attack and defence strategies. We show, first, that
  naive defences don't work against vertex-order attack; second, that
  defences based on simple redundancy don't work much better, but that
  defences based on cliques work well; third, that attacks based on
  centrality work better against clique defences than vertex-order
  attacks do; and fourth, that defences based on complex strategies
  such as delegation plus clique resist centrality attacks better than
  simple clique defences. Our models thus build a bridge between
  network analysis and evolutionary game theory, and provide a
  framework for analysing defence and attack in networks where
  topology matters. They suggest definitions of efficiency of attack
  and defence, and may even explain the evolution of insurgent
  organisations from networks of cells to a more virtual leadership
  that facilitates operations rather than directing them. Finally, we
  draw some conclusions and present possible directions for future
  research.
\end{abstract}

\section{Introduction}

Many modern conflicts turn on connectivity. In conventional war, much
effort is expended on disrupting the other side's command, control and
communications by jamming or destroying his facilities.
Counterterrorism operations involve a similar effort but with
different tools: traffic analysis to trace communications, coupled
with surveillance of the flows of money, material and recruits,
followed by the arrest and interrogation of individuals who appear to
be significant nodes. Terrorists are aware of this, and take measures
to prevent their networks being traced.  Usama bin Laden described his
strategy on the videotape captured in Afghanistan as `Those who were
trained to fly didn't know the others.  One group of people didn't
know the other group' (see~\cite{Krebs}, which describes the
hijackers' networks).

Connectivity matters for social dominance too, as a handful of leading
individuals do much of the work of holding a society together.
Subverting or killing these leaders is likely to be the cheapest way
to make an invaded country submit. When the Norman French invaded
England in the eleventh century, they killed or impoverished most of
the indigenous landowners; when the Turks, and then the Mongols,
invaded India, they killed both landowners and priests; when England
suppressed the Scottish highlands after the 1745 uprising, landowners
were induced to move to Edinburgh or London; and in many of the
dreadful events of the last century, rulers targeted the elite
(Russian kulaks, Polish officers, Tutsi schoolteachers, \ldots).

Moving from politics to commerce, the music industry spends a lot of
money attempting to disrupt peer-to-peer file-sharing networks.
Techniques range from technical attacks to aggressive litigation
against individuals believed to have been running major nodes.

Networks of personal contacts are important in other applications too.
In public health, for example, it often happens that a small number of
individuals account for much of the transmission of a disease. Thus 
Senegal has been more effective at tackling the spread of HIV/AIDS than
other African countries, as they targeted prostitutes~\cite{Thom03}. In
fact, interest in social networks has grown greatly over the last 15
years in the humanities and social sciences~\cite{WF94,CSW2005}.

Recent advances in the theory of networks have provided us with the
mathematical and computational tools to understand such phenomena
better. One striking result is that a network much of whose
connectivity comes from a small number of highly-connected nodes can
be very efficient, but at the cost of extreme vulnerability. As a
simple example, if everyone in the county communicates using one
telephone exchange, and that burns down, then everyone is isolated.

This paper starts to explore the tactical and strategic options open
to combatants in such conflicts. What strategies can one adopt, when
building a network, to provide good trade-offs between efficiency and
resilience? We are particularly interested in complex networks,
involving thousands or millions of nodes, which are so complicated (or
under such dispersed control) that the resilience rules can only be
implemented locally, rather than by a central planner who deliberately
designs a network with multiple redundant backbones.

Is it possible, for example, to create a virtual high-degree node, by
combining a number of nodes which appear on external inspection to
have lower degree? For example, a number of individuals might join
together in a ring, and use some covert communications channel to
route sensitive information round the ring in a manner shielded from
casual external inspection. There is a loose precedent in Chaum's
`dining cryptographers' construction~\cite{Chaum88}, in which a
number of cryptographers pass messages round a ring in such a way as
to mask, from insiders, the source and destination of encrypted
traffic. Can we build a similar construction, but in which the fact
of systematic message routing is concealed from outsiders, with the
result that the participants appear to be `ordinary' nodes making a
modest contribution in the network, rather than important nodes that
should be targeted for close inspection and/or destruction?

\section{Previous Work}

There has been rapid progress in recent years in understanding how
networks can develop organically, how their growth influences their
topology, and how the topology in turn affects both their capacity and
their robustness. There is now a substantial literature: for a
book-length introduction, see Watts~\cite{Watts03}, while literature
surveys are~\cite{AB02,Newmann03}

Early work by Erd\"os and Renyi modelled networks as random
graphs~\cite{ER59,Bollobas}; this is mathematically interesting but
does not model most real-world networks accurately. In real networks,
path lengths are generally shorter; it is well known that any two
people are linked by a chain of maybe half a dozen others who are
pairwise acquainted -- known as the `small-world' phenomenon.  This
idea was popularised by Milgram in the 60s~\cite{Milgram67}.  An
explanation started to emerge in 1998 when Watts and Strogatz produced
the alpha model. Alpha is a parameter that expresses the tendency of
nodes to introduce their neighbours to each other; with $\alpha=0$,
each node is connected to its neighbours' neighbours, so the network
is a set of disconnected cliques, while with $\alpha = \infty$, we
have a random graph. They discovered that, for critical values of
$\alpha$, a small-world network resulted. The alpha model is rather
complex to analyse, so they next introduced the beta network: this is
constructed by arranging nodes in a ring, each node being connected to
its $r$ neighbours on either side, then replacing existing links with
random links according to a parameter $\beta$; for $\beta=0$ no links
are replaced, and for $\beta=1$ all links have been replaced, so that
the network has again become a random graph~\cite{WS98}.  The effect
is to provide a mix of local and long-distance links that models
observed phenomena in social and other networks.

How do networks with short path lengths come about in the real world?
The simplest explanation involves preferential attachment. Barab\'asi
and Albert showed in 1999 how, if new nodes in a network prefer to
attach to nodes that already have many edges, this leads to a
power-law distribution of vertex order which in turn gives rise to a
{\em scale-free} network~\cite{BA99}, which turns out to be a more
common type of network than the alpha or beta types. In a social
network, for example, people who already have many friends are useful
to know, so their friendship is particularly sought by newcomers. In
friendship terms, the rich get richer. There are many economic
contexts in which such dynamics are also of interest~\cite{Jackson03}.

The key paper for our purposes was written by Albert, Jeong, and
Barab\'asi in 2000. They observed that the connectivity of scale-free
networks, which depends on the highly-connected nodes, comes at a
price: the destruction of these nodes will disconnect the network. If
an attacker removes the best-connected nodes one after another, then
past some threshold point the size of the largest component of the
graph collapses~\cite{AJB00}.

Later work by Holme, Kim, Yoon and Han in 2002 extended this from
attacks on vertices to attacks on edges; here, the attacker removes
edges connecting high-degree nodes, and again, past some critical
point, the network becomes disconnected~\cite{Holme02}. They also
suggested using {\em centrality} -- technically, this is the
`betweenness centrality' of Freeman~\cite{F77} -- as an alternative to
degree for attack targeting. (A node's centrality is, roughly
speaking, the proportion of paths on which it lies.) Computing
centrality is harder work for the attacker than observing vertex
degree, but it enables him to attack networks (such as beta networks)
where there is little or no variability in vertex order. Finally, in
2004, Zhao, Park and Lai modelled the circumstances in which a
scale-free network can suffer cascading breakdown from the successive
failure of high-connectivity nodes~\cite{ZPL04}. These ideas find some
resonance in the field of strategic studies: for example, Soviet
doctrine called for destroying a third of the enemy's network, jamming
a further third, and hoping that the remaining third would collapse
under the increased weight of traffic.

\section{Naive Defences Don't Work}

Given the obvious importance of the subject, and the fact that the
Albert-Jeong-Barab\'asi paper appeared in 2000, one obvious question
is why there has been no published work since on how a network can
defend itself against a decapitation attack. Here is one possible
explanation: the two obvious defences don't work.

One of these is simply to replenish destroyed nodes with new nodes,
and furnish them with edges according to the same scale-free rule that
was used to generate the network initially. One might hope that some
equilibrium would be found between attack and defence.

The other obvious defence is to replenish destroyed nodes, but to wire
their edges according to a random graph model. In this way, we might
hope that, under attack, a network would evolve from an efficient
scale-free structure into a less efficient but more resilient random
structure. In a real application, this might happen either as a result
of nodes learning new behaviour, or by selective pressure on a node
population with heterogeneous connectivity preferences: in peacetime
the nodes with higher degree would become hubs, while in wartime they
would be early casualties.

\begin{figure}[hbtp]
\includegraphics[scale=.80]{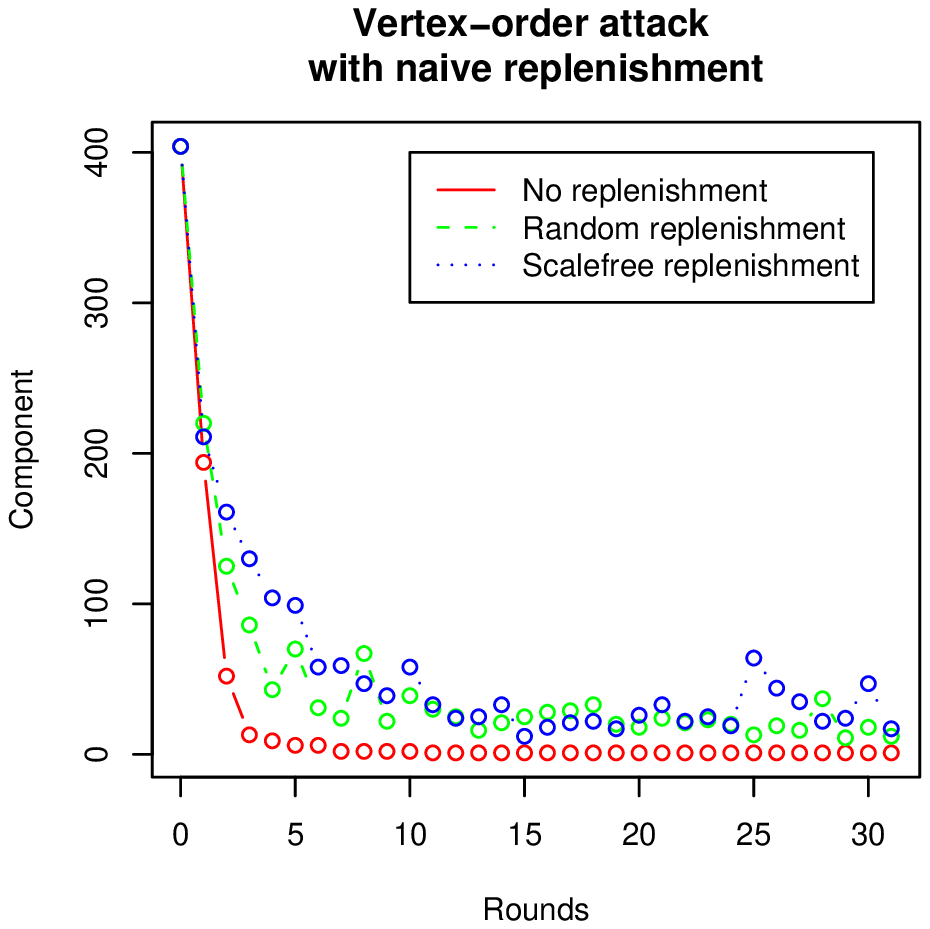}\\
\includegraphics[scale=.80]{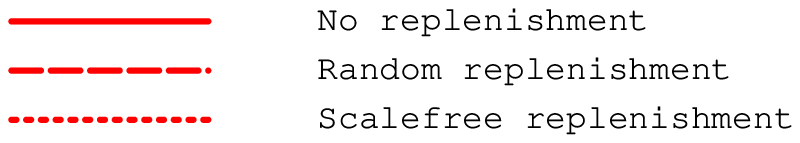}%
\caption{Naive defences against vertex-order decapitation attack}
\label{fig:BA00}
\end{figure}

Nice as these ideas may seem in theory, they do not work at all well
in practice. Figure~\ref{fig:BA00} shows first (solid line) how the
vertex-order attack of Albert, Jeong and Barab\'asi works against a
simulated network with no replenishment, then with random
replenishment, then with scalefree replenishment. In the vanilla case
the attack takes two rounds to disconnect the network; with random
replenishment it takes three, and with scale-free replenishment it
takes four.

It seems that, to defend against these kinds of decapitation attacks
on networks, we will need smarter defence strategies. But how should
these be evolved, and what sort of framework should we use to
evaluate them?

\section{A Model from Evolutionary Game Theory}

Previous researchers considered disruptive attacks on networks to be a
single-round game. Such a model is suitable for applications such as a
conventional war, in which the attacker has to expend a certain amount
of effort to destroy the defender's command, control and
communications, and one wishes to estimate how much; or a single
epidemic in which a certain amount of resource must be spent to bring
the disease under control.

However, there are many applications in which attack and defense
evolve through multiple rounds: terrorism and music-sharing are only
two examples. We now develop a framework for considering this more
general case. We apply ideas from evolutionary game theory developed
by Axelrod and others~\cite{Axel84,Axel97}. This theory studies how
games of multiple rounds differ from single-round games, and it has
turned out to have significant explanatory power in applications from
ethology to economics.

We now formalise a model in which a game is played with a number of
rounds. Each round consists of attack followed by recovery. Recovery
in turn consists of two phases: replenishment and adaptation.

In the {\bf attack phase}, the attacker destroys a number of nodes
(or, in a variant, of edges); this number is his budget. He selects
nodes for destruction according to some rule, which is his strategy.
For example, he might at each round destroy the ten nodes with the
largest number of edges connected to them. He executes this strategy
on the basis of information about the network topology.

In the {\bf replenishment phase}, the defending nodes recruit a
number of new nodes, and go through a phase of establishing
connections -- again, according to given strategies and information.

In the {\bf adaptation} phase, the defending nodes may rewire links
within each connected component of the network, in accordance with
some defensive strategy. The adaptation phase is applied once at the
start of the game, before the first round of attack; thereafter the
game proceeds attack -- replenish -- adapt.

An attack strategy is more efficient, for a given defense strategy, if
an attacker using it requires a smaller budget to disrupt the network.
Similarly, a defense strategy is more efficient if, for a given attack
strategy, it compels the attacker to expend a higher budget to achieve
network disruption. (We will clarify this later once we have presented
and discussed a few simulations.)

We assume initially that the attacker has perfect information about
the network topology, and that her goal is simply to partition the
network -- that is, divide it into two or more nontrivial disjoint
components. We assume that the defender has only local information,
that it, each node shares the information available to those nodes
with which it is connected. Thus, for example, if the attacker manages
to split the network into two components, there is no way for them to
reconnect.  We also start off by assuming that the defence strategy
affects only the adaptation phase, as only once nodes have connected
to a network can they be programmed to follow it; so the replenishment
phase is exogenous.

A further initial assumption is that the attack and defence budgets
are roughly equal. By this we will mean that for each node destroyed
in the attack phase, one node will be replaced in the resource
addition phase. Thus the network will neither grow or shrink in
absolute size and we can concentrate on connectivity effects. We will
discuss other possible assumptions later, but the static budgets and 
global attack / local defence assumptions will get us started.

\section{Defence Evolution -- First Round}

To analyse the vulnerability of a network, the selection of network
elements (nodes or edges) destroyed in each round is the attacker's
choice and constitutes her strategy. The attacker wishes to maximize
the network damage caused per unit of work.

We will start off by considering a static attacker, using what we know
to be a reasonable attack (vertex-order), and examine how the defence
strategy can adapt. Then we will see what better attacks can be found
against the best defence we found. Then we will look for a defence
against the best attack we found in the last round, and so on. There
is no guarantee that the process converges -- there may be a
specialised attack that works well against each defence, and vice
versa -- but if evolutionary games on networks behave like more
traditional evolutionary games, we may expect to find some strategies
that do well overall, as `tit for tat' does in multi-round prisoners'
dilemma. We may also expect to gain useful insights in the process.

\subsection{Defense strategy 1 -- random replenishment}

Our first defensive strategy is the simplest of all, and is one of the
naive defences introduced in the above section. New nodes are joined
to the graph at random. We assume that each attack round removes $r$
nodes, and the replenishment round adds exactly $r$ nodes, each of
which is joined to the surviving vertices with probability $p$. $r$
remains constant for each run of the simulation, while $p$ increases
from $k/(N-r)$ to $k/(N-1)$ as the replenishment proceeds.  In this
strategy, the defender does nothing in the adaptation phase.

This models the case where new recruits to a subversive network simply
contact any other subversives they can find; no attempt is made to
reshape the network in response to the capture of leaders but the
network is simply allowed to become more amorphous.

\subsection{Defense strategy 2 -- dining steganographers}

Our second defensive strategy is more sophisticated, and is inspired
by the theory of anonymous communication as developed by computer
scientists, most notably Chaum~\cite{Chaum88}. A node that acquires a
high vertex order, and thus could be threatened by a vertex-order
attack, splits itself into $n$ nodes, arranged in a ring. The rings
have two functions. First, they provide resilience: a ring broken at
one point still supports communications between all its surviving
nodes, and it is the simplest such structure. Second, nodes can route
covert traffic between appropriate input and output links, and use
encryption and other information-hiding mechanisms to conceal the
traffic. This model was originally presented in Chaum's seminal
`dining cryptographers' paper cited above, so we might refer to it as
the `dining steganographers'. The collaborating nodes in each ring
cannot conceal the existence of communication between them, as the
cover traffic is visible to the attacker. However, from the attacker's
viewpoint it is not obvious that these $n$ nodes are acting as a
virtual supernode.

Our focus here is on the effects of network topology, rather than on
the higher-layer mechanisms that actually implement the covertness
property and that provide any confidentiality of content or of routing
data. We assume a world in which there is sufficient encrypted
traffic (SSL, SSH, DRM, \ldots) that encrypted traffic is not of itself
suspicious so long as it is wrapped in a common ciphertext type. The
attacker's input consists of traffic data collected from the backbone
or from ISPs, and her output consists of decisions to send police
officers to raid the premises associated with particular IP addresses.
Her problem is this: given an observed pattern of communications, whom
should she investigate first?

The precise mechanism of ring formation in our simulation is as
follows. A vulnerable node decides to create a ring and recruits for
the purpose a further $n-1$ nodes from the new nodes introduced in the
most recent replenishment round, or, if they are inadequate, from
among its immediate neighbours. Existing ring members cannot be
recruited, so rings may not overlap. Finally, recruits to a ring
relinquish any existing links with the rest of the network, and the
ring-forming node shares its external links uniformly among all the
members of the ring.

\subsection{Defense strategy 3 -- revolutionary cells}

Our third defensive strategy is inspired by cells of revolutionaries,
along the model favoured historically by a number of insurgent
organisations. A node that acquires a high vertex order splits itself
into $n$ nodes, all linked with each other, with the previous outside
connections split uniformly between them. In graph-theoretic language,
each supernode is a clique.

As in ring formation, a node that considers itself vulnerable is
allowed to split itself into a clique of nodes. The new nodes are
drawn either from the pool of new nodes, or, if they are insufficient,
from low-vertex-order neighbours of the clique-forming node. As
before, this node's external edges are distributed uniformly among
members, while other member nodes' former external edges are deleted.

\subsubsection{Simulations -- first set}

For our first set of simulations, we consider a scalefree network of
$N$ = 400 nodes. We use a Barab\'asi-Albert network created by the
following algorithm:

   \begin{enumerate}
   \item {\em Growth:} Starting with $m_{0}$ = 40 nodes, at each round we
     add $m$ = 10 new nodes, each with 3 edges.

   \item {\em Preferential Attachment:} The probability that a new node
     connects to node $i$
     is $\Pi(k_{i})$ = $k_{i}/\sum_{j}k_{j}$
     where $k_{i}$ is the degree of node $i$.
   \end{enumerate}

Having created the scalefree network, we then ran each of the above
defensive strategies against a vertex-order attack.

\subsubsection{Results}

\begin{figure} [hbtp]
  \includegraphics[scale=.80]{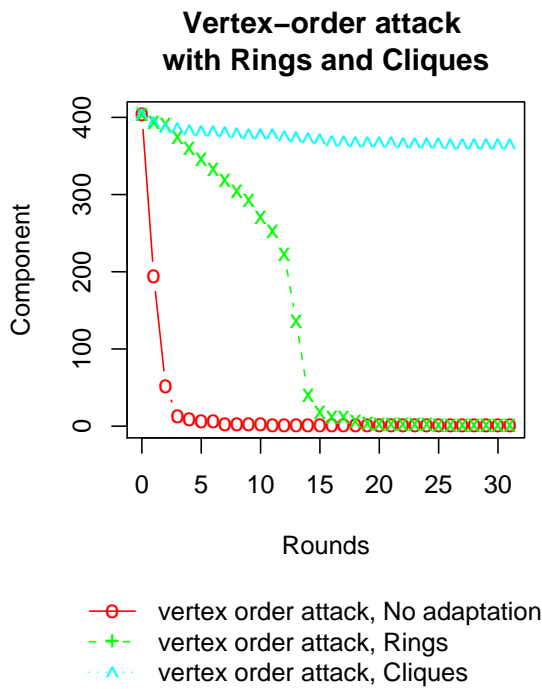}
  \caption{Vertex order decapitation attack in rings, cliques and with
    no adaptation}

    \label{fig:BA01}
\end{figure}

The results of the initial three simulations are given in 
Figure~\ref{fig:BA01}.

The red graph in Figure~\ref{fig:BA01} provides a calibration
baseline. As seen in the above section, random replenishment without
adaptation is ineffective: within three rounds the size of the largest
connected component has fallen by a half, from 400 nodes to well under
200.

The green graph shows that rings give only a surprisingly short-term
defence benefit. They postpone network collapse from about two rounds
qto about a dozen rounds. Thereafter, the network is almost completely
disconnected. In fact, the outcome is even worse than with random
replenishment.

Cliques, on the other hand, work well. A few vertices are disconnected
at each attack round, but as the cyan graph shows, the network itself
remains robustly connected. This may provide some insight into why,
although rings have seemed attractive to theoreticians, those real
revolutionary movements that have left some trace in the history books
have used a cell structure instead.

\section{Attack Evolution -- First Round}

Having tried a number of defence strategies and found that one of
them -- cliques -- is effective, the next step is to try out a number
of attack strategies to see if any of them is effective against our
defences, and in particular against cliques.

Of the attack strategies we tried against a clique defence, the best
performer is an attack based on centrality. We used the centrality
algorithm of Brandes~\cite{Brandes01} to select the highest-centrality
nodes for destruction at each round. As before, our calibration
baseline is random replenishment. For this, the red and black graphs
show performance against vertex-order and centrality attacks
respectively. Both are equally effective; within two or three rounds
the size of the largest connected component has been halved.

The green and blue graphs show that the same holds for rings: the
network collapses completely after about a dozen rounds.  Centrality
attacks are very slightly more effective but there is not much in it.

\begin{figure} [htbp]
   \includegraphics[scale=.80]{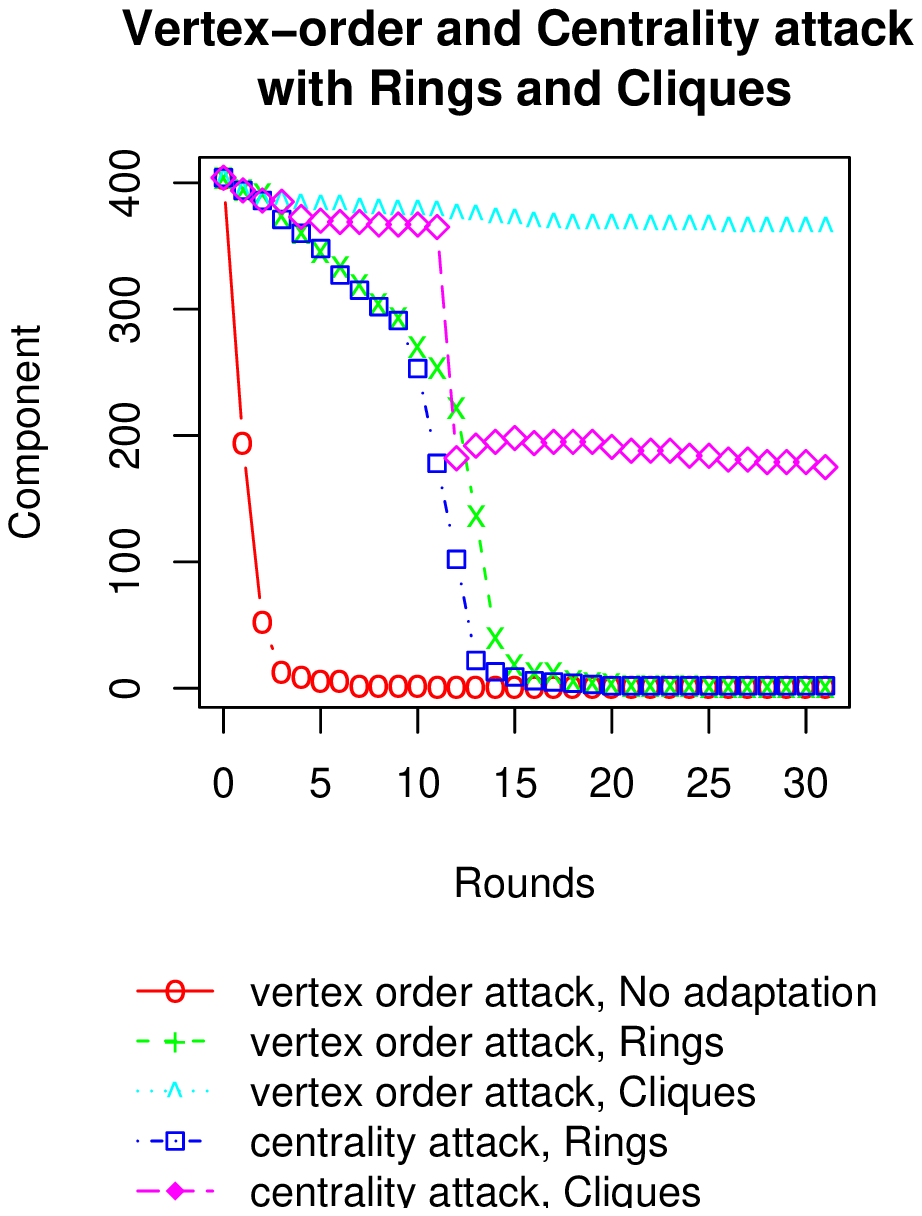}\\%
   \caption{Rings and Cliques defense under vertex order and
     centrality attacks}
    \label{fig:BA02}
\end{figure}

The most interesting results from these simulations come from the
magenta and cyan graphs, which show how cliques behave. Cyan shows, as
before, a vertex-order attack with severity $m$ = 10 being ineffective
against a clique defence. Magenta shows the effect on such a network
of a centrality attack. Here the largest connected component retains
about 400 nodes until the network suddenly partitions at 14 rounds,
whereafter a largest-component size of about 200 is maintained stably.

\begin{figure} [htbp]
   \includegraphics[scale=.80]{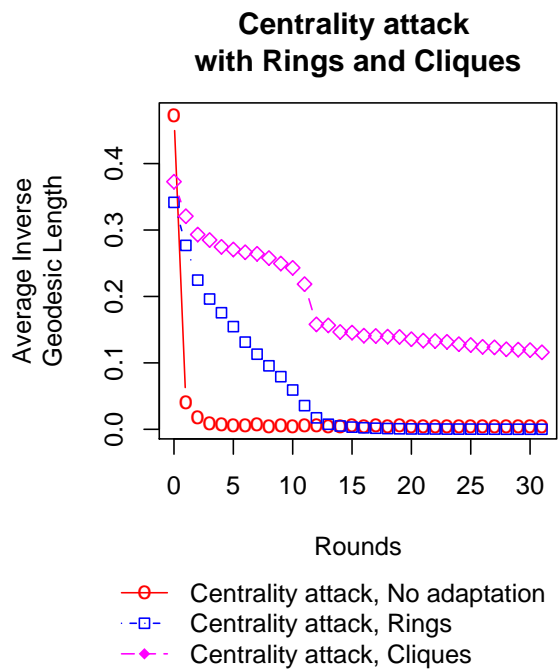}\\%
   \caption{Average inverse geodesic lengths of rings and clique
     adaption, under centrality attack }
    \label{fig:BA02a}
\end{figure}

Some insight into the internal mechanics can be gleaned from
Figure~\ref{fig:BA02a}. This shows the {\em average inverse geodesic
  length}.  For each node, we find the length of the shortest path to
each other node, and take the inverse (we take the length to be
infinite, and thus the inverse to be zero, if the nodes are in
disjoint components).  We average this value over all $n(n-1)/2$ pairs
of nodes. This value falls sharply for defense without adaptation, and
falls steadily for defense with rings.  These falls reflect increasing
difficulty in internode communication.  With cliques, the vertex-order
attack has little effect, while the centrality attack makes steadily
increasing progress on a graph of 400 vertices, until it achieves
partition and reduces the largest component to about 200 vertices. But 
it makes only slow progress thereafter.

\subsection{Clique sizes}

\begin{figure} [htbp]
   \includegraphics[scale=.8]{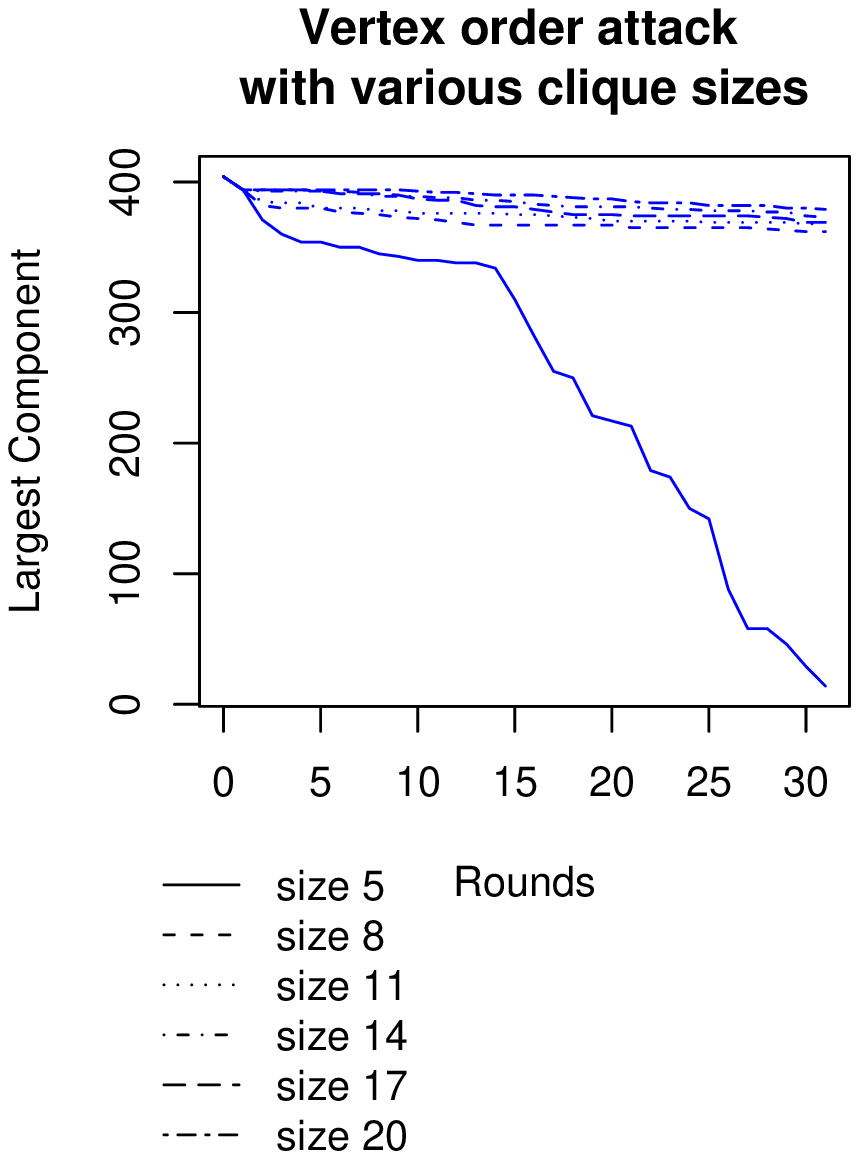}\\%
   \includegraphics[scale=.8]{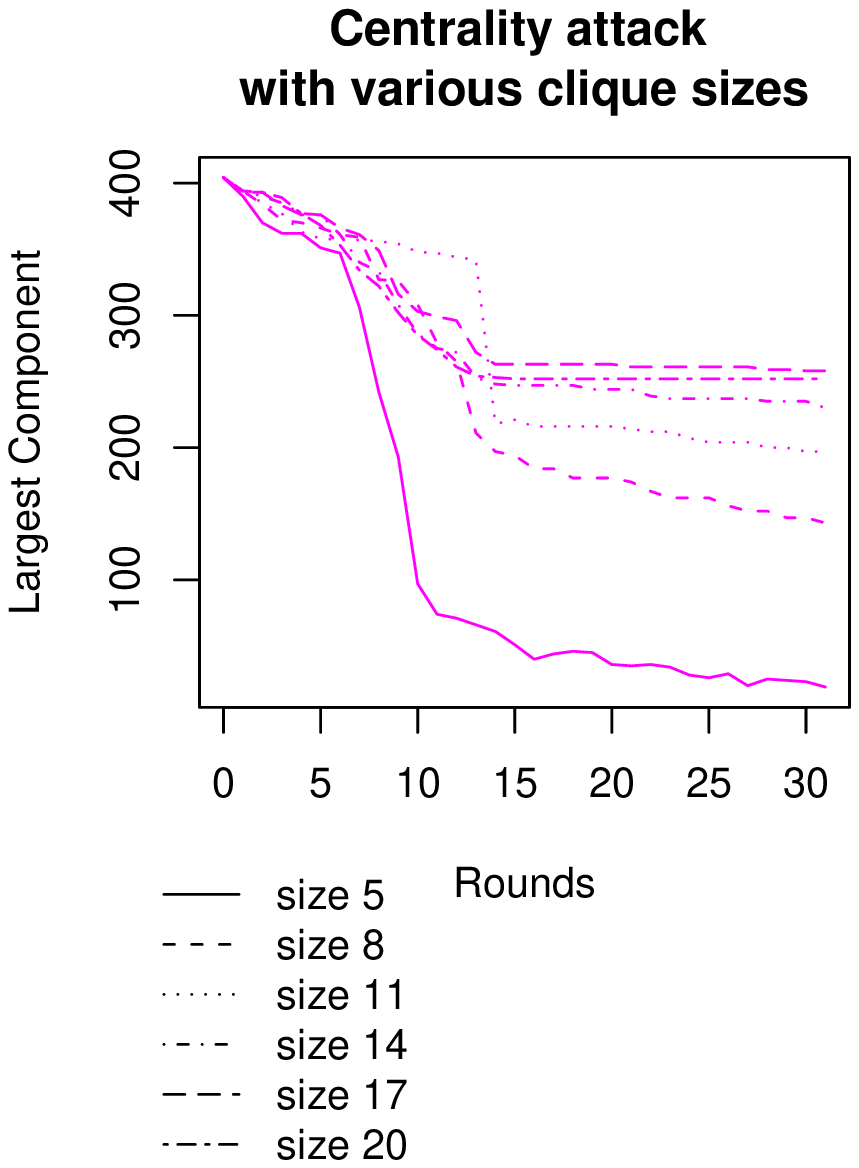}%
   \caption{Clique recovery with different clique sizes under a
     centrality attack  }
    \label{fig:BA03}
\end{figure}

We next ran a simulation comparing how well defense works when using
different sizes of rings and cliques. Ring size appears to make little
difference; rings are just not an effective defence other than in the
very short term. However, varying the clique size yields the results
displayed in Figure~\ref{fig:BA03}.

This shows that under a centrality attack, the performance of the
defense increases steadily with the size of the clique. There is still
a phase transition after about 14 rounds or so after which the largest
connected component becomes significantly smaller, but the size of
this equilibrium component increases steadily from about 150 with
clique size 8 to almost 300 at clique size 20.

\section{Defence Evolution -- Second Round}

Now that we know centrality attacks are powerful, we have tried a
number of other possible defences. The most promising at present
appears to be a compound defence based on cliques and delegation.

The idea behind delegation is fairly simple. A node that is becoming
too well-connected selects one of its neighbours as a `deputy' and
connects it to a second neighbour, with which it then disconnects.
This reflects normal human behaviour even in peacetime: busy leaders
pass new recruits on to colleagues. In wartime, and with an enemy that
might resort to vertex-order attacks, the incentive to delegate is
even greater. Thus a terrorist leader who gets an offer from a wealthy
businessman to finance an attack might simply introduce him to a young
militant who wants to carry one out. The leader need now maintain
communications with at most one of the two.

Delegation on its own is rather slow; it takes dozens of rounds for
delegation to `immunise' a network against vertex-order attack. If a
vanilla scale-free network is going to be exposed to either a
vertex-order or centrality attack from the next round, then drastic
action (such as clique formation) is needed at once; else it will be
disconnected within two or three rounds. Slower defences like
delegation can however play a role, provided they are started from
network formation or a reasonable time period (say 20 rounds) before
the attack begins.

It turns out that the delegation defence, on its own, is rather like
the rings of dining steganographers. Network fragmentation is
postponed (about 14 rounds with the parameters used here) though not
ultimately averted.

\begin{figure} [htbp]
   \includegraphics[scale=.80]{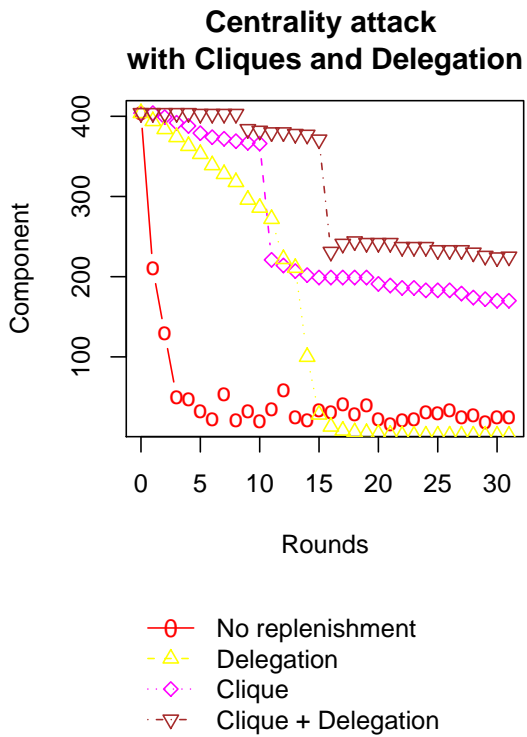}%
   \caption{Component size: clique, immunization by delegation, and
     combined clique and delegation defenses against centrality
     attack}
    \label{fig:BA06}
\end{figure}

What is interesting, however, is this. If we form a network and
immunise it by running the delegation strategy, then run a clique
defence as well from the initiation of hostilities, this compound
strategy works rather better than ordinary cliques. 
Figure~\ref{fig:BA06} shows the simulation results.

\begin{figure} [htbp]
   \includegraphics[scale=.80]{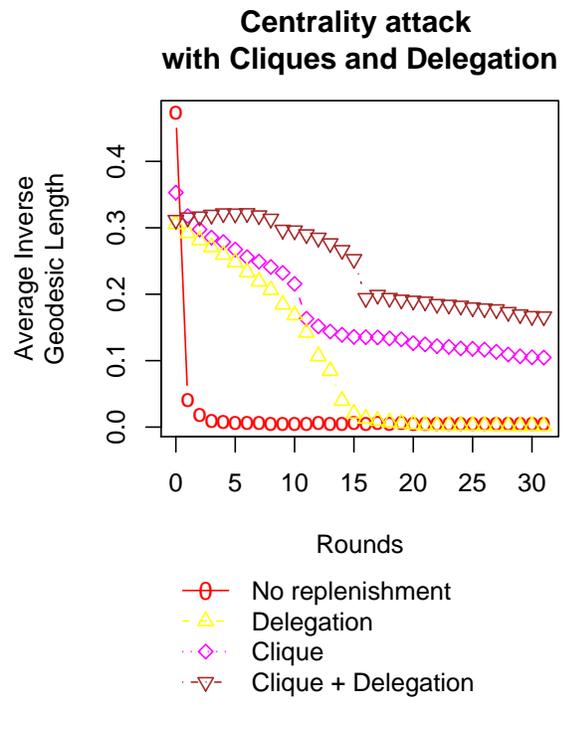}%
   \caption{Clique, immunization by delegation, and combined clique
     and delegation defenses against centrality attack }
    \label{fig:BA07}
\end{figure}

Figure~\ref{fig:BA07} may give some insight into the mechanisms.
Delegation results in shorter path lengths under attack: it
postpones and slows down the growth of path length that otherwise
results from hub elimination. As a result, equilibrium is achieved 
later, and with a larger minimum connected component. 

\section{Conclusions and Future Work}

In this paper, we have built a bridge between network science and
evolutionary game theory.

For some years, people have discussed what sort of communications
topologies might be ideal for covert communication in the presence of
powerful adversaries, and whether network science might be of
practical use in covert conflicts -- whether to insurgents or to
counterinsurgency forces~\cite{BCZ04,Sparrow90}. Our work makes a
start on dealing with this question systematically.

Albert, Jeong and Barab\'asi showed that although a scalefree network
provides better connectivity, this comes at a cost in robustness -- an
opponent can disconnect a network quickly by concentrating its
firepower on well-connected nodes. In this paper, we have asked the
logical next questions. What sort of defence should be planned by
operators of such a network? And what sort of framework can be
developed in which to test successive refinements of attack, defense,
counterattack and so on?

First, we have shown that naive defences don't work. Simply replacing
dead hubs with new recruits does not slow down the attacker much,
regardless of whether link replacement follows a random or scale-free
pattern.

Moving from a single-shot game to a repeated game provides a useful
framework. It enables concepts of evolutionary game theory to be
applied to network problems.

Next, we used the framework to explore two more sophisticated
defensive strategies.  In one, potentially vulnerable high-order nodes
are replaced with rings of nodes, inspired by a standard technique in
anonymous communications.  In the other, they are replaced by cliques,
inspired by the cell structure often used in revolutionary warfare. To
our surprise we found that rings were all but useless, while cliques
are remarkably effective. This may be part of the reason why cell
structures have been widely used by capable insurgent groups.

Next, we searched for attacks that work better against clique
defences.  We found that the centrality attack of Holme et al does
indeed appear to be more powerful, although it can be more difficult
to mount as evaluating node centrality involves knowledge of the
entire topology of the network. Centrality attacks may reflect the
modern reality of counterinsurgency based on pervasive communications
intelligence and, in particular, traffic analysis.

Now we are searching for defences that work better against centrality
attacks. A promising candidate appears to be the delegation defence,
combined with cliques. This combination may in some ways reflect the
reported `virtualisation' strategies of some modern insurgent
networks.

Above all, this work provides a systematic way to evolve and test
security concepts relating to the topology of networks. Clearly the
coevolution of attack and defense can be taken much further. Further
work includes testing:

\begin{enumerate}
\item networks that grow or shrink, maybe with endogenous
  replenishment (current recruitment a function of past operational
  success)
\item imperfectly informed attackers, such as policemen who have
  access to the records of some but not all phone companies or email
  service providers, or who must use purely local measures of
  centrality
\item perfectly informed defenders, who can coordinate connectivity
  globally
\item budget tradeoffs -- for example, a defender might be able to
  hide specific edges but only at some cost to his replenishment
  budget
\item heterogeneous networks, with subpopulations having different
  robustness preferences
\item dynamic strategies that detect opponents' strategies and respond
\item different attacker goals.  For example, some say that the Iraqi
  rebel leader Al-Zarqawi is not bin Laden's subordinate but his
  competitor. So an attack objective might be not just partition, but
  to divide the opposition into groups of less than a certain
  size. When attacking an ad-hoc sensor network, the goal might be to
  reduce the effective bandwidth, and there might be interaction with
  routing algorithms.
\end{enumerate}

Preliminary though it is, we suggest that this work has broad
potential applicability -- from making the Internet more resilient
against natural disasters and malicious attacks, to the question of
how best to disrupt (or design) subversive networks.

\vspace{3ex}

{\large\bf\noindent Acknowledgements:} We have had useful feedback on
an early versionq of this paper from Albert-L\'aszl\'o Barab\'asi, Mike
Bond, George Danezis, Karen Sp\"arck Jones and Chris Lesniewski-Laas.
The first author was supported by a scholarship from the Gates Trust.

\end{document}